# Metal Oxide Nanoparticles and Their Applications: A Report

Namrata Patel[1] and Sandeep Munjal[2]

[1] Shree M M Patel Institute of Science and Research, Gandhinagar, Gujarat, India.

[2] Department of Physics, Indian Institute of Technology Delhi, Hauz Khas, New Delhi-110016, India.

**Abstract**

Herein, we report a brief introduction of metal oxide nanoparticles and their diverse applications in different scientific and medical fields. This report will be updated frequently to give a complete review in similar fields of nanotechnology. In the present version of the report, an introduction to nanotechnology and nanomaterials with some synthesis routes (such as Hydrothermal synthesis and Sol-Gel synthesis etc.) to prepare the metal oxide nanoparticles is given. In this version we have primarily included the basic introduction of application of metal oxide nanoparticles in the fields of biomedical, resistive switching and photovoltaic etc.

1. **<u>Introduction</u>**

Nanotechnology (NT) is an interdisciplinary field that includes nanoscience, nano-electronics, nano-metrology, nano-physics, nano-chemistry, nano-bionics, etc. [1]–[3]. Nanotechnology has emerged as a foundation of a wide range of scientific applications ranging from energy harvesting, to energy storage and to biomedical applications. The concept of nanotechnology was first discussed by renowned physicist Richard Feynman in 1959. Different developing applications of nanotechnology utilize the unique properties arising from the nanoscale dimensions of nanomaterials [4]. The properties of nanomaterials (NMs) can be altered [5] to possess required composition and functionalities, to make them suitable for different application.

The transition-metal oxides (TMOs), which has attracted the attention of materials scientists and engineers, constitute a fascinating class of inorganic solids [6]. Their magnetic [7], electrical [8], optical [9], catalytic [10] and mechanical [11] properties make them technologically useful. TMOs are compounds composed of oxygen atoms bound to transition metals, and are commonly utilized in several electronic applications due to their superior semiconducting properties [12], [13]. TMOs show remarkable physical properties among which spinal oxides (particularly ferrites) show comparably good and interesting magnetic and electrical properties [14], which makes the ferrites particularly very attractive for several novel applications.

In the recent years, the Transition Metal Oxide Nanoparticles (TMO NPs) have emerged as potential candidates for a number of scientific applications including Biomedical Applications [15], Photovoltaic Applications [16], [17], Resistive Switching Applications and many more. Besides physical routes of synthesis of nanoparticles, the chemical routes are preferred because of the availability many easy synthesis techniques such as Hydrothermal and Sol-Gel etc. Further, large scale production and cost-effective synthesis of nanomaterials is possible through chemical synthesis to match industry requirements.

2. **Synthesis**

Here is a brief description of Hydrothermal and Sol-Gel technique for the synthesis of nanoparticles -

**2.1.** Hydrothermal Synthesis

Hydrothermal/solvothermal methods are low temperature (~ 70 - 250°C), and high pressures synthesis methods, in which the materials are synthesized in the presence of water/polar inorganic solvents [18]. The reactions take place in closed containers kept above the boiling point of the used polar solvent. As a result, the developed high pressure decreases the boiling point of the solvent as well as increases the solubility of the precursors, which provides better condition for the growth of crystals. The autoclaves, which are used in the hydrothermal synthesis, are chemically resistant stainless steel (SS) vessels. These SS autoclaves are available in numerous capacities/sizes ranging from few ml to few litres. Teflon (polytetrafluoroethylene) lined cups are used to perform the hydrothermal reaction. Generally, the Teflon cup is filled about 50-70 % of its maximum capacity with the final reaction solution. This Teflon cup is then transferred to the SS autoclave, which is then tightly closed and placed into an oven at required temperature. After completion of the reaction, the obtained nanoparticles are washed and can be used in several applications.

**2.2.** Sol-Gel Synthesis

Sol-gel synthesis is a chemical solution process [19], [20] to synthesize the ceramic and glass materials in the form of thin films, fibres, or powders. A sol is a colloidal (the dispersed phase is so small that gravitational forces do not exist; only Van der Waals forces and surface charges are present) or molecular suspension of solid particles of ions in a solvent. A gel is a semi-rigid mass that forms when the solvent from the sol begins to evaporate and the particles or ions left behind begin to join together in a continuous network [21]. After a drying process, a thermal treatment (calcination) is performed in order to favour further crystallization of the material.

**3. Applications**

3.1. Biomedical Applications

MO NPs such as $Fe_3O_4$, $NiFe_2O_4$ and $CoFe_2O_4$ have proven their potential in several biomedical applications from targeted drug delivery to magnetic hyperthermia applications [22]. For such application, it is necessary that the NPs to be used must be biocompatible. Besides biocompatibility, water dispersibility is also an essential requirement for these NPs to be used in many biomedical applications such as Magnetic Hyperthermia applications. By careful reduction of size of magnetic nanoparticles, they can be turned into superparamagnetic nanoparticles and can be useful for practical applications such as magnetic hyperthermia. This water dispersibility can be achieved by coating several ligands on the surface of the nanoparticles [23], [24]. Possible biomedical applications of magnetic nanoparticles are depicted in the figure given below -

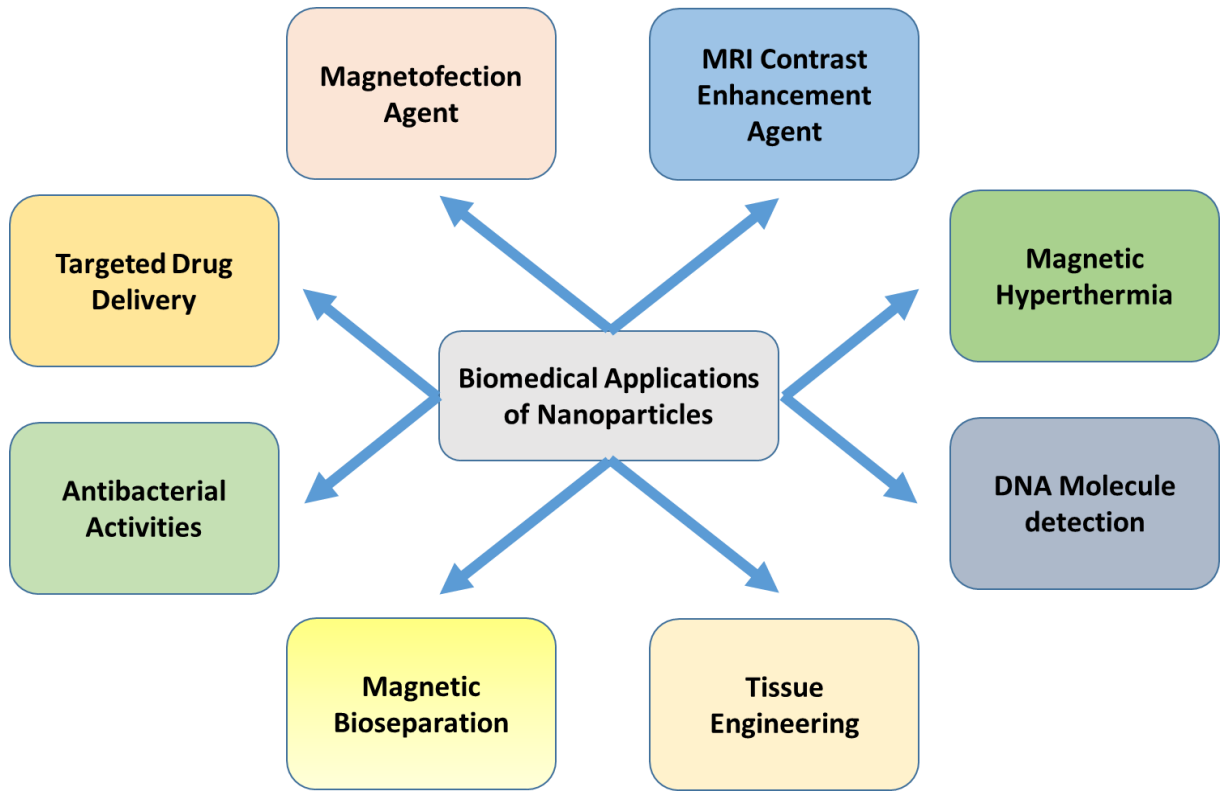

**Figure 1** Biomedical applications of nanoparticles.

3.2. Photovoltaic Applications

Semiconducting materials can absorb solar energy and utilize this energy for the transition of electrons from the valence band to the conduction band. This transition leads to create an electron hole pair in the semiconductor materials, and this phenomenon is used to generate the electric energy from the solar energy. Many metal oxides such as ZnO, $TiO_2$ etc. have good semiconducting properties with a wide band gap. These metal oxides can easily be synthesized in the form of nanoparticles and these nanoparticles can further be used to fabricate solar cells [25].

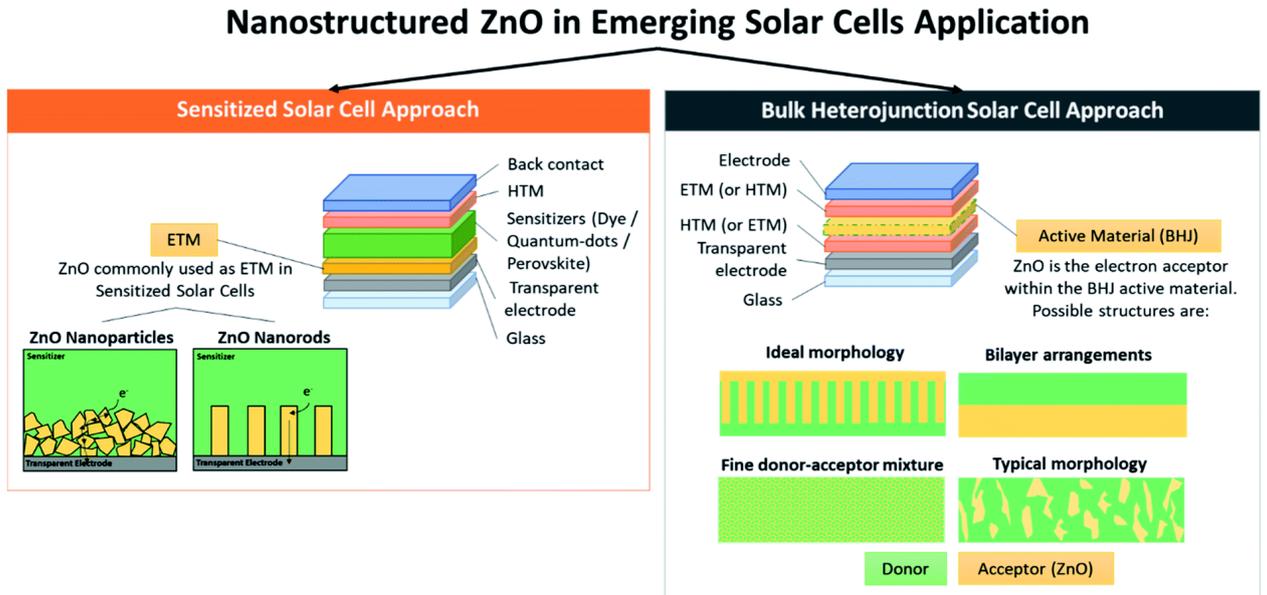

**Figure 2** Nanostructured ZnO in emerging solar cell applications [25].

3.3. Resistive Switching

The Resistive Random Access Memory (ReRAM) is one of the most potential emerging non-volatile memory is based on resistive switching (RS) phenomena. In RS phenomena the electrical resistance of an insulating/dielectric layer sandwiched between two metal electrodes is altered by the application of an electric field [26]–[29]. For many ReRAM devices initially an ''Electroforming'' process is usually required before the memory cell starts switching between different resistance states [2], [26]. In this process the resistive switching device is forced by a higher electric field to develop a conducting filament between top and bottom electrodes [30]. This electroforming process is generally not desirable and in order to get reproducible and uniform resistive switching the it is desired to develop electroforming free resistive switching devices. By using oxide nanoparticles in the resistive switching active layer, we can obtain the electroforming free resistive switching devices [30]. **Table 1** lists the resistive switching (RS) memory devices fabricated using transition metal oxide nanoparticles or their composites as an active layer.

**Table 1:** Resistive switching devices fabricated using transition metal oxide nanoparticles or their composites as an active layer.

| S. No. | Active Layer Material | Reference |
|---|---|---|
| 1 | $NiFe_2O_4$ nanoparticles | [31] |
| 2 | Iron oxide (IO) nanoparticles (NPs) embedded in polyvinyl alcohol (PVA) | [32] |
| 3 | $CoFe_2O_4$ nanoparticles | [33] |
| 4 | $TiO_2$ nanoparticles | [34] |
| 5 | $CoFe_2O_4$ nanoparticles | [30] |
| 6 | MnO nanoparticles | [35] |
| 7 | Carbon-$TiO_2$ Nanoparticles | [36] |
| 8 | PMMA/Al:ZnO Composite Films | [37] |

## 4. Conclusion

In summary we have provided a brief detail on the synthesis and application of transition metal oxide nanoparticles. Sol-Gel and Hydrothermal techniques for the synthesis have been discussed and some applications of metal oxide nanoparticles such as resistive switching and photovoltaic have also been discussed.

First draft:2013

This version: 2021